\documentclass[10pt,conference,letterpaper]{IEEEtran}
\IEEEoverridecommandlockouts

\usepackage[a4paper, total={184mm,239mm}]{geometry}
\def\BibTeX{{\rm B\kern-.05em{\sc i\kern-.025em b}\kern-.08em
    T\kern-.1667em\lower.7ex\hbox{E}\kern-.125emX}}

\makeatletter
\newcommand\notsotiny{\@setfontsize\notsotiny\@vipt\@viipt}
\makeatother

\usepackage{graphicx}
\usepackage[hidelinks]{hyperref}
\usepackage{amsmath}
\usepackage{amssymb}
\usepackage{cite}
\usepackage{url}
\usepackage{booktabs}
\usepackage{amsfonts}
\usepackage{longtable}
\usepackage{nicefrac}
\usepackage{microtype}
\usepackage{xcolor}
\usepackage{enumitem}
\usepackage{comment}
\usepackage{subcaption}
\usepackage[most]{tcolorbox}
\tcbuselibrary{skins, breakable}
\usepackage{ulem}
\usepackage{listings}
\usepackage{pdfpages}
\usepackage{multirow}
\usepackage{verbatim}
\usepackage{tabularx}
\usepackage{pifont}
\usepackage{lipsum}
\usepackage{colortbl}
\usepackage{float}
\usepackage{hyphenat}
\usepackage{array}
\usepackage{wrapfig}
\usepackage{fancyvrb}

\newcommand*\colourcheck[1]{%
  \expandafter\newcommand\csname #1check\endcsname{\textcolor{#1}{\ding{51}}}
}
\colourcheck{blue}
\colourcheck{green}
\colourcheck{red}
\colourcheck{teal}
\newcommand*\colourcross[1]{%
  \expandafter\newcommand\csname #1cross\endcsname{\textcolor{#1}{\ding{55}}}%
}
\colourcross{blue}
\colourcross{green}
\colourcross{red}
\colourcross{purple}
\colourcross{gold}

\usepackage{algorithmic}
\usepackage{textcomp}

\newtcolorbox{casebox}[1]{
    colback=white,
    colframe=blue!50!black,
    fonttitle=\bfseries,
    fontupper=\scriptsize, 
    title=#1,
    enhanced,
    drop shadow={black!50!white},
    breakable,
    boxsep=2pt,        
    left=2pt, right=2pt, 
    top=2pt, bottom=2pt,
    before skip=5pt,
    after skip=5pt
}

\lstdefinestyle{verilog}{
    basicstyle=\ttfamily\tiny,
    keywordstyle=\color{blue},
    commentstyle=\color{green!60!black},
    stringstyle=\color{red},
    numbers=left,
    numberstyle=\tiny\color{gray},
    stepnumber=1,
    numbersep=5pt,
    backgroundcolor=\color{white},
    frame=single,
    rulecolor=\color{black!30},
    tabsize=4,
    captionpos=b,
    breaklines=true,
    breakatwhitespace=true,
    showstringspaces=false
    aboveskip=0pt,     
    belowskip=0pt,     
    abovecaptionskip=2pt,
    belowcaptionskip=2pt,
}

\input{preamble/preamblelistingsconf}
\newcommand{\ignore}[1]{{}}

\newcommand{\squishlist}{
	\begin{list}{$\bullet$}
		{ \setlength{\itemsep}{0pt}
			\setlength{\parsep}{1pt}
			\setlength{\topsep}{1pt}
			\setlength{\partopsep}{0pt}
			\setlength{\leftmargin}{0.9em}
			\setlength{\labelwidth}{1.5em}
			\setlength{\labelsep}{0.4em} } }
	\newcommand{\squishend}{
	\end{list}  }

\definecolor{graphFirst}{RGB}{2,136,209} 
\definecolor{graphSecond}{RGB}{211,47,47} 
\definecolor{graphThird}{RGB}{245,124,0} 
\definecolor{graphFourth}{RGB}{56,142,60} 
\definecolor{graphFifth}{RGB}{81,45,168} 
\definecolor{graphSixth}{RGB}{69,90,100} 
\definecolor{graphSeventh}{RGB}{251,192,45} 
\definecolor{backgroundSecond}{RGB}{239,154,154} 
\definecolor{backgroundThird}{RGB}{255,204,128} 
\definecolor{backgroundFourth}{RGB}{165,214,167} 
\definecolor{backgroundFifth}{RGB}{179,157,219} 
\definecolor{backgroundSixth}{RGB}{176,190,197} 
\definecolor{backgroundSeventh}{RGB}{255,245,157} 

\newcommand{\circleone}{\ding{202}}
\newcommand{\circletwo}{\ding{203}}
\newcommand{\circlethree}{\ding{204}}
\newcommand{\circlefour}{\ding{205}}
\newcommand{\circlefive}{\ding{206}}
\newcommand{\circlesix}{\ding{207}}
\newcommand{\circleseven}{\ding{208}}
\newcommand{\circleeight}{\ding{209}}
\newcommand{\circlenine}{\ding{210}}

\usepackage{tabularx}

\author{%
    \IEEEauthorblockN{Jitendra Bhandari\IEEEauthorrefmark{1}, Vineet Bhat\IEEEauthorrefmark{1}, 
    Yuheng He\IEEEauthorrefmark{2},   
    Hamed Rahmani\IEEEauthorrefmark{1},
    Siddharth Garg\IEEEauthorrefmark{1}
    and Ramesh Karri\IEEEauthorrefmark{1}
    }
    \IEEEauthorblockA{%
    \IEEEauthorrefmark{1}New York University Tandon School of Engineering
    \IEEEauthorrefmark{2}Cornell University }
  \vspace{-2ex}
}

\begin{document}

\title{Masala-CHAI: A Large-Scale SPICE Netlist Dataset for Analog \underline{C}ircuits by \underline{H}arnessing \underline{AI}}


\maketitle

\IEEEpubidadjcol

\begin{abstract}

Masala-CHAI is a fully automated framework leveraging large language models (LLMs) to generate Simulation Programs with Integrated Circuit Emphasis (SPICE) netlists. It addresses a long-standing challenge in circuit design automation: automating netlist generation for analog circuits. Automating this workflow could accelerate the creation of fine-tuned LLMs for analog circuit design and verification. In this work, we identify key challenges in automated netlist generation and evaluate multimodal capabilities of state-of-the-art LLMs, particularly GPT-4, in addressing them. We propose a three-step workflow to overcome existing limitations: labeling analog circuits, prompt tuning, and netlist verification. This approach enables end-to-end SPICE netlist generation from circuit schematic images, tackling the persistent challenge of accurate netlist generation. We utilize Masala-CHAI to collect a corpus of 7,500 schematics that span varying complexities in 10 textbooks and benchmark various open source and proprietary LLMs. Models fine-tuned on Masala-CHAI when used in LLM-agentic frameworks such as AnalogCoder achieve a notable 46\% improvement in Pass@1 scores. We open-source our dataset and code for community-driven development.


\end{abstract}


\section{Introduction}

Large Language Models (LLMs) have received significant attention due to their wide-ranging applications, from text summarization to code generation, and have a growing impact across various fields. For hardware design, LLMs have primarily demonstrated potential in the digital domain. 
This includes tasks such as Verilog code generation~\cite{liu2023verilogeval,thakur2023autochip,lu2024rtllm,thakur2023verigen}, assertion generation~\cite{kande2023llmassisted,fang2024assertllm}, bug fixing~\cite{qiu2024autobench,bhandari2024llm}, and electronic design automation (EDA) tool scripting~\cite{wu2024chateda,liu2023chipnemo}.
The success of these domain-tailored LLMs relies on access to large and high-quality datasets. For example, the Verigen code generation model~\cite{thakur2023verigen} was trained on a dataset of 75,000 Verilog files sourced from GitHub, with additional data extracted from Verilog textbooks to further enhance its performance in Verilog-related tasks.
Beyond hardware design, textbooks have also been valuable in other domains, such as systems biology~\cite{sayed2018recipes} and protein interaction studies~\cite{holtzapple2020flute}, demonstrating their broader utility in training domain-specific models.

\begin{figure}[thbp]
\vspace*{-0.1in}
\centering

\hspace*{-0.1in}
\includegraphics[width=\columnwidth]{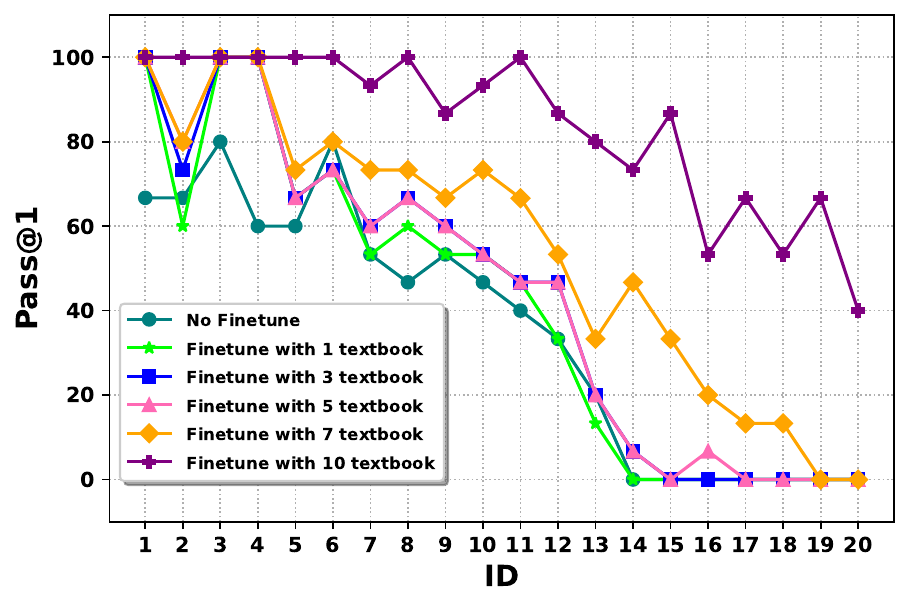}
\caption{Pass@1 performance of GPT-4o fine-tuned with Masala CHAI datasets extracted from between 1-10 textbooks, compared to the no fine-tuning baseline (task IDs from Table II). Our largest dataset of ~7500 captioned SPICE netlists extracted from 10 textbooks provides more than 40\% Pass@1 accuracy even on challenging tasks like telescoping cascode amplifier and bandgap reference circuit generation.}
\label{fig:motivate}
\vspace*{-0.2in}
\end{figure}
 
Building on the success of LLMs in the digital domain, it is natural to extend their application in the \textit{analog domain}, particularly in the automated generation of analog circuits from natural language specifications. Analog circuits
are described in SPICE and its various variants, which serve as the industry-standard textual representation for analog circuit simulation. SPICE provides a low-level description that defines the interconnections between analog components like resistors, capacitors, inductors, and transistors.   
The availability of open-source datasets for SPICE remains significantly limited compared to Verilog, which poses a major challenge for automating the learning of frameworks for the generation of net lists~\cite{tao2024amsnet,lai2024analogcoder,llama3modelcard,img2sim,img2sim2}. However, analog circuit textbooks and research papers are plentiful and contain a trove of analog circuit diagrams, 
but these are usually in \emph{image} (or figure) format. These images must then be manually converted to SPICE netlists, which is painstaking and time-consuming. 


\textbf{Masala-CHAI}\footnote{Masala is the Hindi word for ``SPICE"} is an \emph{automated} framework
for SPICE netlist generation from analog circuit schematics. We incorporate various techniques, including (i) Schematic extractor from documents, (ii) fine-tuned circuit component detectors to extract passive and active components, (iii) deep hough transform priors to identify nets, (iv) extensive prompt tuning for LLMs to fix common errors in netlist extraction, and (v) post-extraction verification. We curated the largest open-source analog circuit dataset, comprising 7,500 examples using Masala-CHAI, and fine-tuned various LLMs, including \texttt{CodeLlama}, \texttt{DeepSeek-34B}, and \texttt{GPT-(3.5,4)}. As shown in ~\autoref{fig:motivate}, larger datasets significantly enhance performance, improving \texttt{pass@k} results across different numbers of textbooks used for fine-tuning.


Our contributions are as follows: 
\begin{enumerate}
[noitemsep,nolistsep,leftmargin=*]
\item  An empirical case study to understand the limitations of state-of-the-art multi-modal LLMs like GPT-4o (shown in AMSNet~\cite{tao2024amsnet}) in SPICE netlist extraction from schematics.  
\item  Build \textbf{Masala-CHAI}, the \emph{first fully automated framework that combines LLMs with custom-trained deep network models} for large-scale SPICE netlist extraction from circuit schematics found in textbooks and research papers. 
\item Leverage Masala-CHAI to compile an open-sourced dataset of $\sim$7,500 SPICE netlists with metadata (figure captions, components, and nodes) from 10 textbooks. To the best of our knowledge, this is 
the \emph{largest} curated and labeled SPICE dataset available currently. 
\item Finetune and benchmark four state-of-art LLMs (CodeLLama-70B, DeepSeek-34B, GPT-3.5, GPT-4o) on the Masala-CHAI dataset, showcasing its potential for automated SPICE netlist generation from English prompts.
\item Enhance the AnalogCoder~\cite{lai2024analogcoder} agentic framework by incorporating GPT-4o fine-tuned on Masala-CHAI, achieving a 46$\%$ absolute improvement in Pass@1 scores.

\end{enumerate}


\section{Related Work}
LLMs have gained significant attention in chip design~\cite{zhong2023llm4eda}. Tremendous progress has been made in improving Verilog code generation~\cite{liu2023verilogeval,thakur2023autochip,lu2024rtllm,thakur2023verigen}. These advances demonstrate how LLMs can streamline and improve digital hardware design workflows. In addition to code generation, prompt engineering has proven effective in chip design~\cite{Blocklove_2023,fu2023gpt4aigchip,chang2023chipgpt}. Using LLMs, researchers have efficiently conceptualized and designed complex digital hardware architectures, facilitating faster and more accurate chip development. Beyond code generation, LLMs have found applications in assistive chatbots, script generation, and bug analysis~\cite{liu2023chipnemo}. ~\cite{wu2024chateda} explores LLMs in planning and executing tasks in the Electronic Design Automation (EDA) flow.
The use of LLMs in generating assertions and testbenches for verifying the correctness of Integrated Circuit (IC) designs has also seen notable improvements~\cite{kande2023llmassisted,fang2024assertllm,bhandari2024llm,qiu2024autobench}. 
~\cite{img2sim,img2sim2} have explored schematic-to-netlist generation using ML-based approaches. However, these solutions are not reproducible due to the unavailability of any open-source implementation.
Recent advances extended LLM applications into analog design. Closed-loop Python code generation for analog circuits by LLM agents has shown promising results~\cite{lai2024analogcoder}. LLMs are able to generate circuit topologies from  specifications~\cite{chang2024lamagic}. A dataset for exploring SPICE netlist generation was recently released in~\cite{tao2024amsnet}, summarized in~\autoref{tab:amsnet}. 

\begin{table}[tbh]
\centering

\caption{Comparison with AMSNet~\cite{tao2024amsnet}. Here, `M' denotes Manual, `P' denotes Partial Automation (with some manual intervention), and `A' denotes Fully Automated.}
\label{tab:amsnet}
\resizebox{\columnwidth}{!}{

\begin{tabular}{ccccccc}
\toprule
\multirow{2}{*}{ } & \multirow{2}{*}{ Schematic } & \multirow{2}{*}{ Component } & \multirow{2}{*}{ Nets } &  \multirow{2}{*}{ Textbooks } & \multirow{2}{*}{ Dataset } \\
& & & & &  \\ \midrule
AMSNet~\cite{tao2024amsnet} & M & \textbf{A} & P &  1 & 734  \\
\textbf{Masala-CHAI} & \textbf{A} & \textbf{A} & \textbf{A} & 10 & \textbf{7500} \\
\bottomrule
\end{tabular}
}
\end{table}

\section{Key Challenges in SPICE Netlist Extraction\label{sec:challenge}}

Despite advances in multi-modal LLMs, models like GPT-4o do not automatically extract accurate SPICE netlists from circuit schematics~\cite{tao2024amsnet}.
Our work begins with an in-depth analysis of GPT-4o's failure modes. Building on these insights, we propose our solutions and introduce a fully automated SPICE netlist extraction framework without manual annotations.

\subsection{Can GPT-4o detect electrical components accurately?}

\begin{figure}[thbp]
\vspace*{-0.1in}
\centering

\hspace*{-0.1in}
\includegraphics[width=\columnwidth]{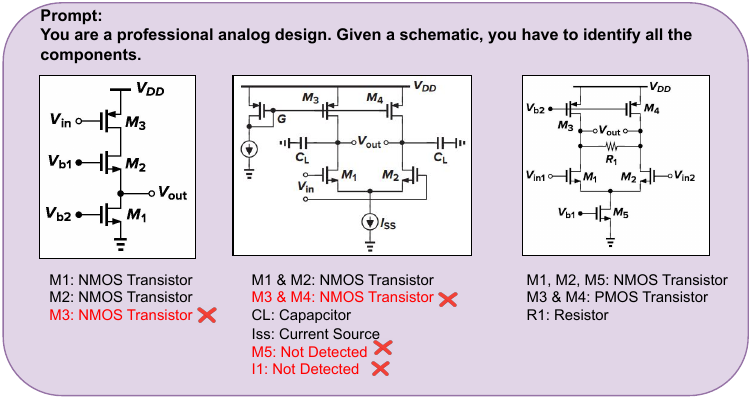}
\caption{GPT-4o responses when asked to list all components in three sample circuit schematics.}
\label{fig:kc1}
\end{figure}

\begin{figure}[tb]
    \centering
\begin{tcolorbox}[width=1.0\linewidth, halign=left, colframe=black, colback=white, boxsep=0.01mm, arc=1.5mm, left=2mm, right=2mm, boxrule=0.5pt]\footnotesize
You are an expert analog designer. You will be provided with a schematic, your task is to follow the below instructions carefully: 
\begin{enumerate}
    \item To identify the NMOS and PMOS MOSFET, follow the instructions carefully. For NMOS, the arrow on the source terminal points outwards from the transistor. For PMOS, the arrow on the source terminal points inward towards the transistor.
    \item List all the components correctly.
   
\end{enumerate} 
\end{tcolorbox}
\caption{Prompt Tuning guides the GPT to differentiate between NMOS and PMOS.}
\label{fig:prompt1}
\vspace*{-0.2in}
\end{figure}

When generating a SPICE netlist, accurately identifying all circuit components is essential for designers and challenging for language models. We curated a set of analog schematics featuring resistors, capacitors, inductors, MOSFETs, and various sources from a popular analog textbook~\cite{razavi2000design} and evaluated GPT-4o’s ability to recognize electrical components in circuit schematics. As shown in \autoref{fig:kc1}, GPT-4o frequently misidentifies components, often confusing NMOS and PMOS transistors (left and middle schematics) or omitting key elements like PMOS transistors and current sources. However, in some cases (final schematic), it accurately identifies all components, demonstrating potential despite its inconsistencies.

\textbf{\textit{Solution 1:}} \textit{Customized Object Detection Network for Analog Circuit Components-} 
Recognizing the limitations of GPT-4o in this task, we propose a dedicated solution using a deep CNN-based object detection model, using the strong performance of models such as YOLO~\cite{Jocher_Ultralytics_YOLO_2023}. By training a CNN model designed \emph{explicitly} for schematic components, we significantly improve the accuracy in detecting and bounding all circuit elements, a method detailed in more detail in Section~\ref{sec:framework}.

\textbf{\textit{Solution 2:}} \textit{Prompt Tuning -} 
We developed a targeted prompt enhancement strategy to refine LLM component differentiation, especially between NMOS and PMOS transistors. By emphasizing structural differences in prompts (see \autoref{fig:prompt1}), we achieved notable improvements in accuracy. Our empirical findings support this approach in Section~\ref{sec:framework}.

\subsection{Can GPT-4o accurately connect circuit components?}

\begin{figure}[h]
\vspace*{-0.15in}
\centering
\includegraphics[width=\columnwidth]{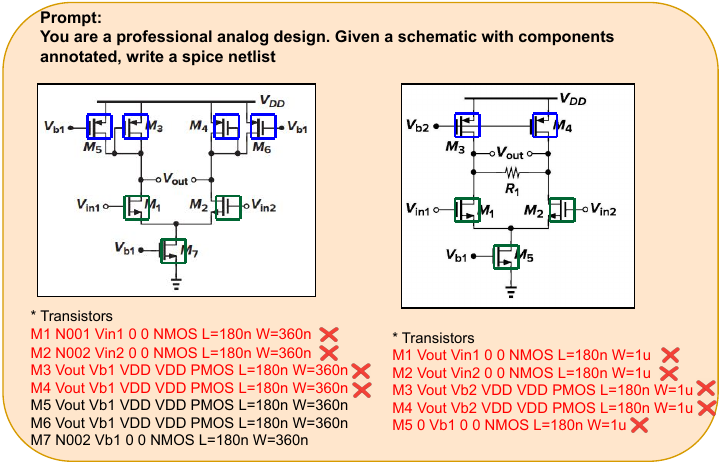}
\caption{SPICE netlist generated by the GPT-4o. For brevity, we only show the part of the netlist that describes transistors.}
\label{fig:kc2}
\vspace*{-0.2in}
\end{figure}

After detecting all circuit components, the next step is to ensure their correct connections, so the generated netlist accurately represents the schematic. While GPT-4o successfully maps 2-terminal devices, our study uncovered several critical failure modes that compromise the accuracy of netlist generation. These failure modes are illustrated in~\autoref{fig:kc2}.
Firstly, GPT-4o incorrectly assumes that intersecting nets are connected, even when no connection exists. This error arises partly from inconsistencies in the definition of connectivity across different schematic notations.
Secondly, GPT-4o often misidentifies MOSFET terminals, confusing the drain, gate, and source. Despite the drain and source's electrical equivalence, this misidentification affects the accuracy of the final netlist.
Finally, differential input and output voltage pairs pose significant challenges for GPT-4o. For example, in the left schematic of \autoref{fig:kc2}, while components were correctly identified, the source connections for (M1 and M2) were incorrect, and the gate voltage for (M3 and M4) was wrongly assumed to be `Vb1'. Similar errors appeared in the right schematic, where differential output was mishandled, and M5's drain was incorrectly mapped.

\textbf{\textit{Solution:}} \textit{Automatic Net Annotation using Deep Hough Transform -} 
Leveraging our knowledge of component locations, we developed a method to identify and annotating nets using a Deep Hough transform line priors. Incorporating these annotations into the prompt (see Section~\ref{sec:framework}), led to improvement in GPT-4o's performance in translating schematics to SPICE netlists. This systematic annotation reduced ambiguities and enhanced the LLM's ability to interpret and connect components, particularly in complex schematics correctly.



\section{Methodology \label{sec:framework}}

\begin{figure*}[htbp]
\vspace*{-0.02in}
\centering
\includegraphics[width=\textwidth,height=2.3in]{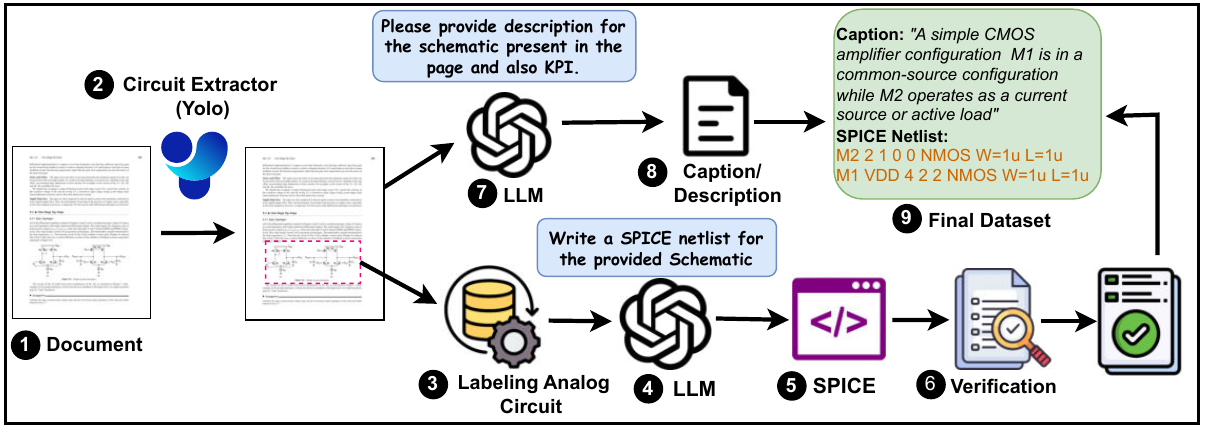}
\vspace*{-0.02in}
\caption{Dataset preparation flow starting from a PDF document consisting of Schematic and related text. Masala-CHAI extracts the relevant information and generates the SPICE netlist for the schematic.}
\label{fig:flow}

\end{figure*}

\begin{figure*}[h]
\centering
\includegraphics[width=\textwidth,height=3.4cm]{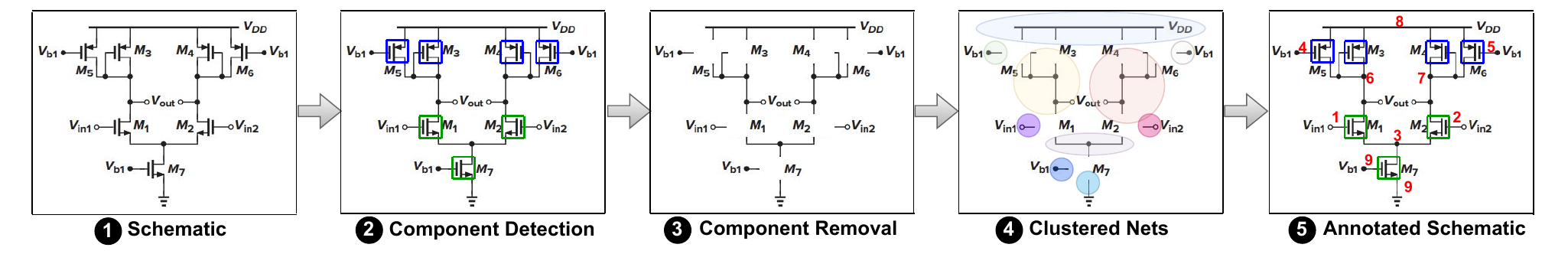}
\vspace*{-0.1in}
\caption{Schematic $\longrightarrow$ Annotated Schematic. Colors denote different component types. Similarly, different nets are clustered.}
\label{fig:preprocess}
\vspace*{-0.2in}
\end{figure*}

This section details the methodology used, addressing the challenges outlined in Section~\ref{sec:challenge}. Accurate generation of SPICE netlists remains a bottleneck for LLMs due to difficulties in correctly identifying circuit components and establishing net connections.
Our dataset creation framework, shown in~\autoref{fig:flow}, starts by paginating a textbook into individual PDFs (\circleone{}), which are then passed through a 
YOLOv8~\cite{Jocher_Ultralytics_YOLO_2023},
a state-of-the-art object localization and classification model, to extract schematics from the page.
Along with extracted schematics, we also extract and summarize text describing the schematic, for example, from the schematic caption or text on the page referring to the schematic. 
These are used as additional context for fine-tuning LLMs, Section~\ref{sec:finetune}, shown in~\autoref{fig:flow} (\circleseven{}  - \circlenine{}). 
The extracted schematic is still in image format and needs to be reliably converted to a SPICE netlist.
To this end, \textbf{Masala-CHAI} follows a three-step process: (i) labeling analog circuits, (ii) prompt tuning, and (iii) SPICE netlist verification, starting from schematic extraction (~\autoref{fig:flow}, \circlethree{} - \circlesix{}). These steps are described next.

\subsection{Labeling Analog Circuit}
~\autoref{fig:flow} (\circlethree{}) simplifies the LLM's task of generating SPICE netlists by detecting components and facilitating net annotation. As discussed in Section~\ref{sec:challenge}, SPICE netlist generation is not straightforward, requiring the designer to feed information alongside the schematics. However, this manual process hinders generating large-scale datasets. Thus, we automate the labeling of all parts of the schematics.

\subsubsection{Detection of Circuit Components}

We train a custom YoloV8 model~\cite{Jocher_Ultralytics_YOLO_2023} specifically for detecting and classifying circuit components with high precision, 
similar to prior work~\cite{img2sim,img2sim2}. YoloV8's Darknet53 backbone~\cite{bochkovskiy2020yolov4}, enhanced with a ``Cross Stage Partial" bottleneck~\cite{wang2020cspnet}, integrates high-level semantic information with low-level spatial details, enabling accurate detection even for small components---a requirement for component detection in complex circuit schematics. 
We fine-tune YOLOv8 on an open-source dataset~\cite{v3qwe_dataset} consisting of ~4,300 circuit diagrams with annotated bounding boxes across 12 component classes: \texttt{AC Source, BJT, Battery, Capacitor, DC Source, Diode, Ground, Inductor, MOSFET, Resistor, Current Source, and Voltage Source}. Images were resized to \texttt{640x640} and trained for \texttt{1000}  epochs with a learning rate of \texttt{0.01}.
The resulting fine-tuned YoloV8 model localizes each component in the schematic by outputting its center coordinates, a bounding box, and classification labels,
as illustrated in~\autoref{fig:preprocess} (\circletwo{}).

\subsubsection{Net Detection}

We performed net detection in circuit schematics by leveraging pre-trained deep hough transform line priors~\cite{lin2020ht}, a state-of-the-art approach for line detection in images. Unlike conventional methods, this model operates in the Hough domain, parametrizing line segments in polar coordinates to achieve precise line segmentation. We remove components from the schematic to address challenges posed by line segments within components such as capacitors and MOSFETs, leaving only the lines representing nets, as illustrated in \autoref{fig:preprocess} (\circlethree{}). This ensures a clean input for subsequent net analysis and eliminates potential sources of ambiguity.

Following net detection, we propose a simple yet effective heuristic for clustering line segments into nets. Specifically, we group all line segments into a single cluster if their endpoints fall within a radius of 40 pixels, as demonstrated in \autoref{fig:preprocess} (\circlefour{}). To ensure reliability, we tune the radius parameter and manually verify the consistency of the clustered nets. 

\subsection{Prompt Tuning}

Prompts are crucial in guiding LLMs to generate precise, high-quality outputs. Crafting effective prompts involves meticulous design of their language, structure, and contextual framing, enabling models to respond accurately and meaningfully to user inputs. Through prompt refinement, designers can precisely control model behavior, ensuring the generation of coherent and domain-specific responses. 
In this study, we systematically explore diverse prompt designs tailored to enhance understanding of schematics. Our approach demonstrates significant effectiveness, as illustrated in \autoref{fig:prompt1} and \autoref{fig:prompt2}. Specifically, prompt in \autoref{fig:prompt2} is used at the stage~\autoref{fig:flow} (\circlefour{}), facilitating insightful queries to the GPT model.

\begin{figure}[h]
    \centering
\begin{tcolorbox}[width=1.0\linewidth, halign=left, colframe=black, colback=white, boxsep=0.01mm, arc=1.5mm, left=2mm, right=2mm, boxrule=0.5pt]\footnotesize
You are an expert analog designer. You will be provided with a schematic, your task is to follow the below instructions carefully: 
\begin{enumerate}
    \item List all components which you can observe from the figure.
    \item MOSFETs are 3 terminal devices with (drain, gate, source).
    \item For each component, look at the net number in red.
    \item To identify source terminal of a MOSFET, choose net highlighted in red which is nearest to the arrow of the MOSFET.
    \item Write a SPICE netlist.
   \end{enumerate} 
\end{tcolorbox}
\caption{Prompt GPT to generate SPICE netlist from Schematic.}
\label{fig:prompt2}

\end{figure}

\subsection{SPICE netlist verification}

Our case study revealed that some generated netlists (\autoref{fig:flow} (\circlefive{})) still exhibit issues such as floating nets, posing risks
large-scale datasets where manual inspection is infeasible.  
To mitigate this, we introduced a final verification step (\autoref{fig:flow} (\circlesix{})) that calls an  
LLM on the extracted SPICE netlist to automatically identify and correct common errors, for instance, removing floating nets as mentioned above. These checks are provided in our open-source code base, and can be further extended to address other sources of error, as required.
\section{Results\label{sec:results}}

\subsection{Dataset Creation}

\begin{figure*}[thbp]
\vspace*{-0.1in}
\centering

\hspace*{-0.2in}
\includegraphics[width=\textwidth, height=1.8in]{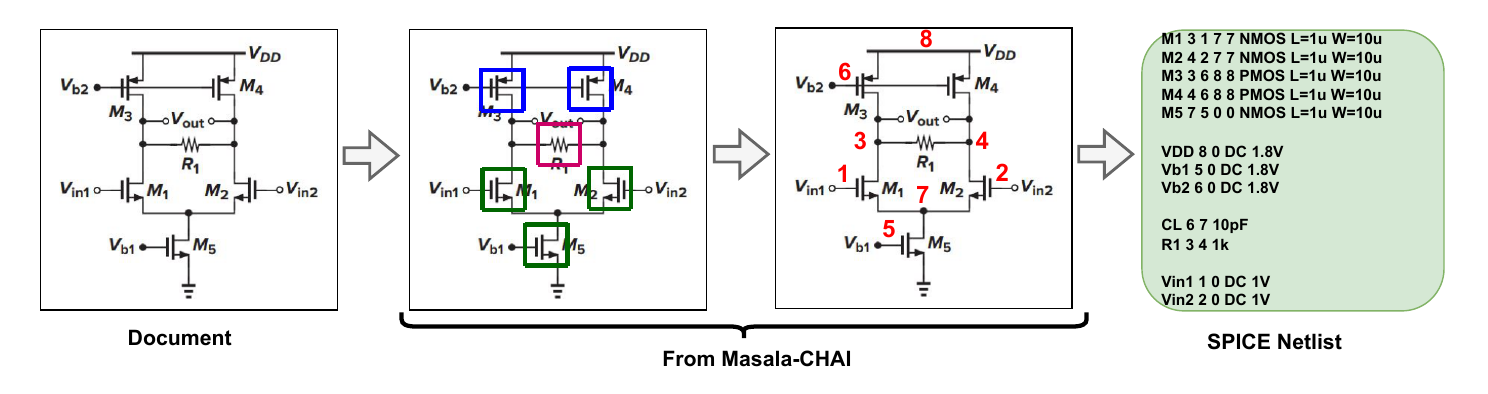}
\caption{Automatic SPICE Netlist generation using Masala-CHAI.}
\label{fig:dataset1}
\vspace*{-0.2in}
\end{figure*}

\begin{figure}[t]
    \centering
    \subfloat[\label{fig:comp}]{
            \includegraphics[width=0.48\columnwidth]{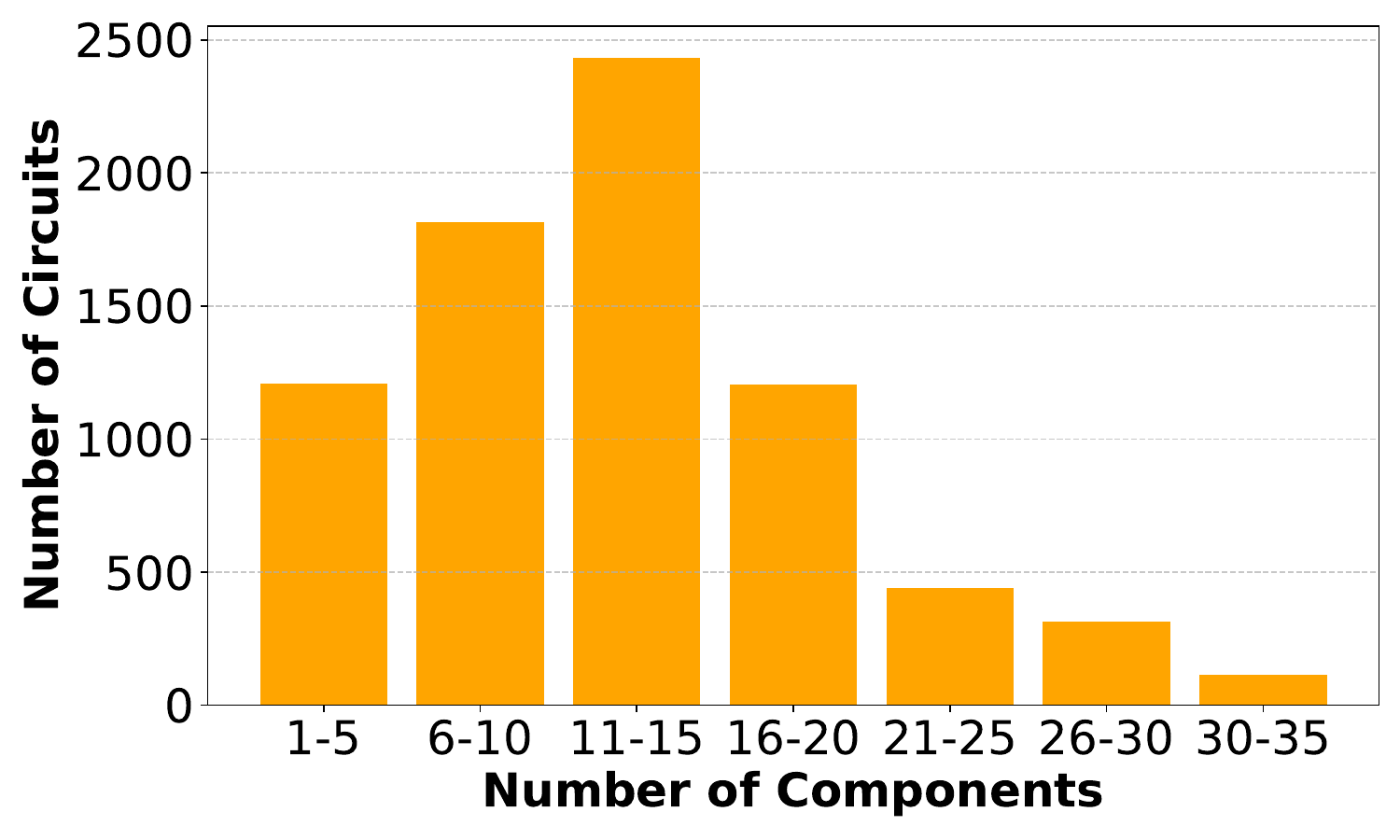}
    }
    \subfloat[\label{fig:node}]{
        {\includegraphics[width=0.48\columnwidth]{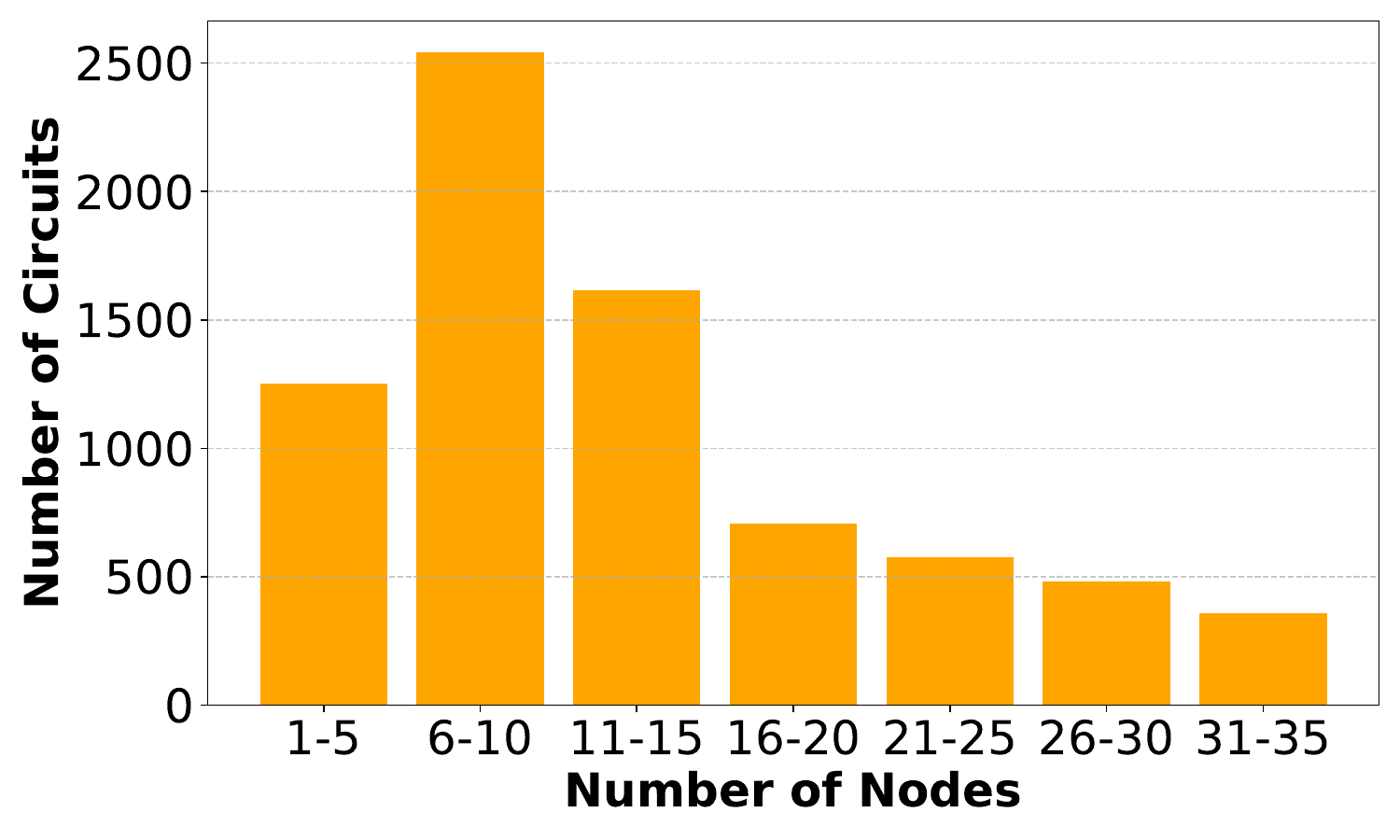}}
    }
    \\
     \subfloat[\label{fig:mosfet}]{
            \includegraphics[width=0.48\columnwidth]{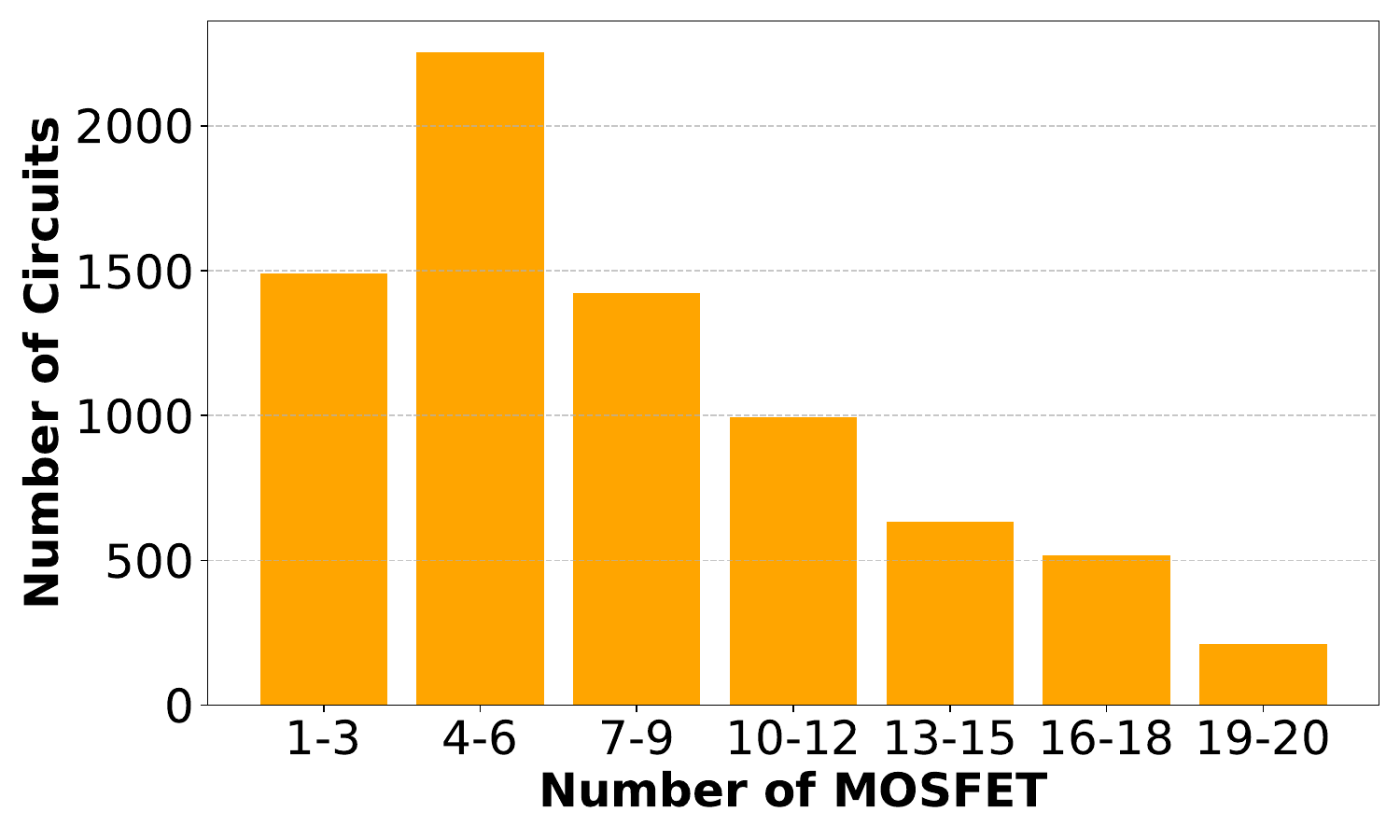}
    }
    \subfloat[\label{fig:lines}]{
        {\includegraphics[width=0.48\columnwidth]{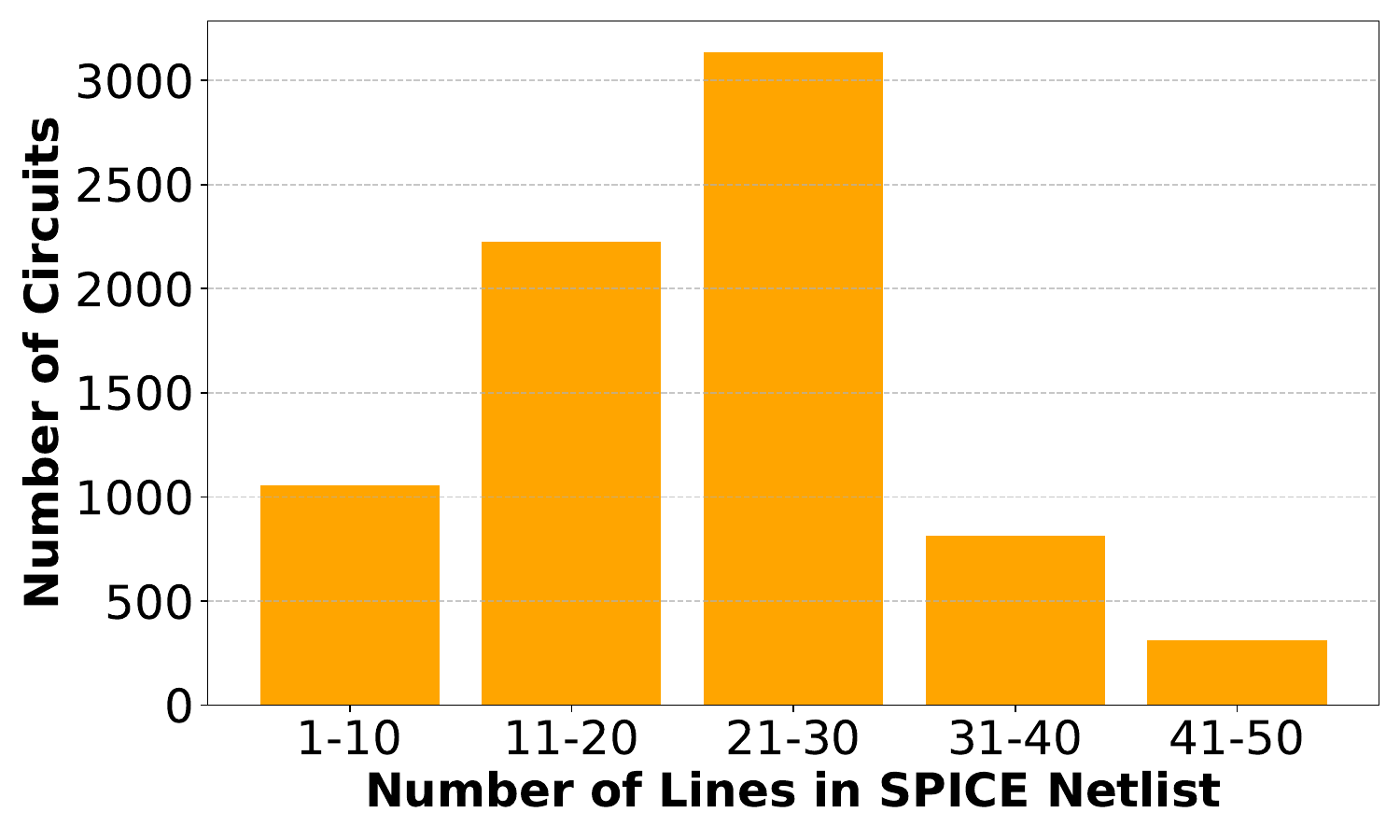}}
    }
    \caption{Distribution of (a) \# Components, (b) \# Nodes, (c) \# MOSFETs, and (d) \#lines of SPICE code in the dataset.
    \label{fig:distribution} }
    \vspace*{-0.2in}
\end{figure}

We curated a comprehensive dataset by automating schematic extraction and SPICE netlist generation from 10 textbooks~\cite{book_1,book_2,book_3,book_4,book_5,book_6,book_7,book_8,book_9,book_10}. These textbooks were selected for their high-quality schematic images, clear context, and detailed explanations.
After collecting $\sim$7,500 schematic images, we applied the Masala-CHAI framework (\autoref{fig:flow}) to generate SPICE netlists. Alongside, we collected captions and descriptions as detailed in Section~\ref{sec:framework}.~\autoref{fig:dataset1} demonstrates how Masala-CHAI translates a schematic into SPICE. Since our goal is to generate structurally correct netlists, each component is marked with default parameters.

Our dataset is characterized to highlight its diversity and complexity in ~\autoref{fig:distribution}: (a) shows the variation in the number of components within schematics, where a higher count indicates greater circuit complexity; (b) presents the distribution of nodes, representing component connectivity. Accurately identifying and mapping MOSFET terminals to the correct nets is a significant challenge for LLMs; (c) focuses on the number of MOSFETs, where an increase in count corresponds to higher complexity in SPICE netlist generation; Finally, (d) illustrates the distribution of SPICE netlist line counts, emphasizing the circuit complexities. This characterization provides insights into the dataset, aiding in the evaluation of automated SPICE generation techniques.
By curating this diverse and complex dataset, we establish a robust benchmark for evaluating and improving LLM performance in automated SPICE netlist generation. For more analysis see Appendix~\ref{sec:appendix_a}
and ~\ref{sec:appendix_b}.

\begin{table*}[h]
\centering

\caption{Evaluation benchmarks. We refer each design by its ID.}
\label{tab:benchmark}
\resizebox{\textwidth}{!}{
\begin{tabular}{cl|cl|cl}
\toprule
ID & Circuit & ID & Circuit & ID & Circuit \\
\midrule
1 & Common-source amplifier & 
2 & 2-stage common src amplifier w/ res. load &
3 & Common-drain amplifier
\\
4 & Common-gate amplifier  &
5 & Single-Stage RC Low-Pass Filter &
6 & Src Degenerated Amplifier \\
7 &  Current Mirror & 
8 & Common-source amplifier using active load &  
9 & Cascode amplifier using NMOS and res. load \\
10 & 1-stage differential amplifier &
11 & Diode-connected Amplifier & 
12 & Buffer design using MOSFET \\ 
13 & 2-input NAND gate &
14&2-stage op-amp with Miller compensation& 
15 & SRAM cell with 6 transistors \\
16 & 2-stage op-amp with diff. i/p and single-handled o/p  &
17 & Fully Diff. Amplifier w/ Common-Mode Feedback &
18 & Cross-coupled LC oscillator \\
19 & Telescopic cascode op amplifier &
20 & Bandgap Reference Amplifier &&\\
\bottomrule
\end{tabular}
}

\end{table*}


\begin{table*}[tbh]
\centering

\caption{Results on LLMs and Finetuned (`FT') variants on MASALA-CHAI dataset. `CL' is CodeLlama. `DS' is DeepSeek.}
\label{tab:result}
\resizebox{\textwidth}{!}{
\begin{tabular}{ccccccccccccccccc}
\toprule
\multirow{2}{*}{ ID } & \multicolumn{2}{c}{CL-70B} & \multicolumn{2}{c}{CL-70B+FT} & \multicolumn{2}{c}{DS-34B } & \multicolumn{2}{c}{DS-34B + FT} & \multicolumn{2}{c}{GPT-3.5} & \multicolumn{2}{c}{GPT-3.5 + FT} & \multicolumn{2}{c}{GPT-4o } & \multicolumn{2}{c}{GPT-4o + FT} \\ \cmidrule(lr){2-3} \cmidrule(lr){4-5} \cmidrule(lr){6-7} \cmidrule(lr){8-9} \cmidrule(lr){10-11} \cmidrule(lr){12-13}
\cmidrule(lr){14-15} \cmidrule(lr){16-17}
Pass& @1 & @5 & @1 & @5 & @1 & @5 & @1 & @5 & @1 & @5 & @1 & @5 & @1 & @5 & @1 & @5 \\
\midrule
1 & 13.3 & 57.1 & 26.7 & 84.6 & 26.7 & 84.6 & 40 & 95.8 & 66.7 & \textbf{100} & 93.3 & \textbf{100}& 66.7 & \textbf{100} & \textbf{100} & \textbf{100} \\

2 & 0.0 & 0.0 & 13.3 & 57.1 & 6.7 & 33.3 & 26.7 & 84.6 & 53.3 & 99.3 & 86.7 & \textbf{100} & 66.7 & \textbf{100} & \textbf{100} & \textbf{100} \\

3 & 26.7 & 84.6 & 33.3 & 91.6 & 13.3 & 57.1 & 46.7 & 98.1 & 66.7 & \textbf{100} & \textbf{100}& \textbf{100} & 80 & \textbf{100} & \textbf{100} & \textbf{100} \\

4 & 13.3 & 57.1 & 33.3 & 91.6 & 26.7 & 84.6 & 33.3 & 91.6 & 60 & 99.8 & 80 & \textbf{100} & 60 & 99.8 & \textbf{100} & \textbf{100} \\

5 & 6.7 & 33.3 & 26.7 & 84.6 & 6.7 & 33.3 & 26.7 & 84.6 & 46.7 & 98.1 & 86.7 &\textbf{100} & 60 & 99.8 & \textbf{100} & \textbf{100} \\

6 & 0.0 & 0.0 & 0.0 & 0.0 & 0.0 & 0.0 & 13.3 & 57.1 & 60.0 & 99.8 & 73.3 & \textbf{100} & 80 & \textbf{100} & \textbf{100} & \textbf{100} \\

7 & 0.0 & 0.0 & 6.7 & 33.3 & 0.0 & 0.0 & 26.7 & 84.6 & 46.7 & 98.1 & 86.7 & \textbf{100} & 53.3 & 99.3 & \textbf{93.3} & \textbf{100} \\

8 & 0.0 & 0.0 & 0.0 & 0.0 & 0.0 & 0.0 & 6.7 & 33.3 & 40.0 & 95.8 & 66.7 & \textbf{100} & 46.7 & 98.1 & \textbf{100} & \textbf{100} \\

9 & 0.0 & 0.0 & 0.0 & 0.0 & 0.0 & 0.0 & 0.0 & 0.0 & 46.7 & 98.1 & 73.3 & \textbf{100} & 53.3 & 99.3 & \textbf{86.7} & \textbf{100}\\

10 & 0.0 & 0.0 & 0.0 & 0.0 & 0.0 & 0.0 & 0.0 & 0.0 & 26.7 & 84.6 & 86.7 & \textbf{100} & 46.7 & 98.1 & \textbf{93.3} & \textbf{100} \\

11 & 0.0 & 0.0 & 0.0 & 0.0 & 0.0 & 0.0 & 0.0 & 0.0 & 0.0 & 0.0 & 66.7 & \textbf{100} & 40 & 95.8 & \textbf{100} & \textbf{100} \\

12 & 0.0 & 0.0 & 0.0 & 0.0 & 0.0 & 0.0 & 0.0 & 0.0 & 13.3 & 57.1 & 66.7 & \textbf{100} & 33.3 & 91.6 & \textbf{86.7} & \textbf{100} \\

13 & 0.0 & 0.0 & 0.0 & 0.0 & 0.0 & 0.0 & 0.0 & 0.0 & 0.0 & 0.0 & 53.3 & 99.3 & 20.0 & 73.6 & \textbf{80} & \textbf{100} \\

14 & 0.0 & 0.0 & 0.0 & 0.0 & 0.0 & 0.0 & 0.0 & 0.0 & 0.0 & 0.0 & 40 & 95.8 & 0.0 & 0.0 & \textbf{73.3} & \textbf{100} \\

15 & 0.0 & 0.0 & 0.0 & 0.0 & 0.0 & 0.0 & 0.0 & 0.0 & 0.0 & 0.0 & 33.3 & 91.6 & 0.0 & 0.0 & \textbf{86.7} & \textbf{100} \\

16 & 0.0 & 0.0 & 0.0 & 0.0 & 0.0 & 0.0 & 0.0 & 0.0 & 0.0 & 0.0 & 40 & 95.8 & 0.0 & 0.0 & \textbf{53.3} & \textbf{99.3} \\

17 & 0.0 & 0.0 & 0.0 & 0.0 & 0.0 & 0.0 & 0.0 & 0.0 & 0.0 & 0.0 & 20 & 73.6 & 0.0 & 0.0 & \textbf{66.7} & \textbf{100} \\

18 & 0.0 & 0.0 & 0.0 & 0.0 & 0.0 & 0.0 & 0.0 & 0.0 & 0.0 & 0.0 & 0 & 0.0 & 0.0 & 0.0 & \textbf{53.3} & \textbf{99.3} \\

19 & 0.0 & 0.0 & 0.0 & 0.0 & 0.0 & 0.0 & 0.0 & 0.0 & 0.0 & 0.0 & 13.3 & 57.1 & 0.0 & 0.0 & \textbf{66.7} & \textbf{100} \\

20 & 0.0 & 0.0 & 0.0 & 0.0 & 0.0 & 0.0 & 0.0 & 0.0 & 0.0 & 0.0 & 0 & 0.0 & 0.0 & 0.0 & \textbf{40} & \textbf{95.8} \\
\bottomrule
\end{tabular}

}
\end{table*}

\subsection{Finetuning}
\label{sec:finetune}

Fine-tuning LLMs is widely used to adapt models to specific downstream tasks~\cite{thakur2023verigen,lu2024rtllm}. The model leverages knowledge acquired from pre-training on large unsupervised language learning tasks, allowing its weights to be adjusted using a smaller, task-specific dataset. This enables the model to capture novel patterns and improve task-specific reasoning without training from scratch. Fine-tuning LLMs offers several key advantages:
(i) Reduced training time and computational costs by leveraging pre-trained foundational knowledge,
(ii) Enhanced performance on downstream tasks, particularly beneficial when data is scarce, and
(iii) Improved transfer learning across subtasks within a domain, increasing efficiency in model deployment.


\textbf{Benchmarking LLMs:} We evaluated the capabilities of LLMs in analog circuit design, using both open-source (\texttt{CodeLlama-70B}~\cite{codellama}, \texttt{DeepSeek-V2}~\cite{liu2024deepseek}) and proprietary (\texttt{GPT-3.5,4o-mini}~\cite{openai2024gpt4}) models. We fine-tuned all baseline LLMs using the corpus collected through the MASALA-CHAI framework. Open-source models were fine-tuned on a Nvidia A100 80GB GPU utilizing LoRA~\cite{hu2022lora}, while GPT models were fine-tuned using their respective fine-tuning playground~\cite{finetune}.

\textbf{Metrics:}
We adopt the \emph{Pass@k} metric, widely used in code-generation tasks~\cite{thakur2023verigen,lu2024rtllm,lai2024analogcoder}, as our main evaluation measure. We define $\text{Pass@k} = 1 - \frac{\binom{n - c}{k}}{\binom{n}{k}}$ (for $k \in \{1,5\}$), where $c$ is the number of successful trials out of $n$ total trials. In our experiments, we set $n=15$ for all the scenarios.

\textbf{Testing:} We selected 20 design problems as shown in ~\autoref{tab:benchmark}, frequently used as standalone circuits or subcomponents in larger designs, to form our evaluation benchmark. LLMs were prompted to generate the corresponding SPICE netlists.

\textbf{Discussion:}~\autoref{tab:result} summarizes the \texttt{Pass@k} performance of various LLMs for SPICE netlist generation. As anticipated, open-source models (\texttt{DeepSeek} and \texttt{CodeLlama}) underperform relative to the proprietary \texttt{GPT} models, both before and after fine-tuning. Nonetheless, fine-tuning on the collected dataset, (ID=1--5) provides a slight increase in the \texttt{Pass@k} for these open-source models, enabling coverage for additional design IDs (ID=6--8). Similarly, fine-tuned \texttt{GPT-3.5} and \texttt{GPT-4o} achieve higher scores (ID=1--13) and non-zero performance for design IDs (ID=14--20). Notably, the fine-tuned \texttt{GPT-4o} (\textit{``Best Result”}) successfully generates SPICE netlists for all design IDs, highlighting the efficacy of the \textbf{MASALA-CHAI} framework (also illustrated in~\autoref{fig:motivate}). 

\begin{figure}[h]
\vspace*{-0.1in}
\centering
\includegraphics[width=\columnwidth]{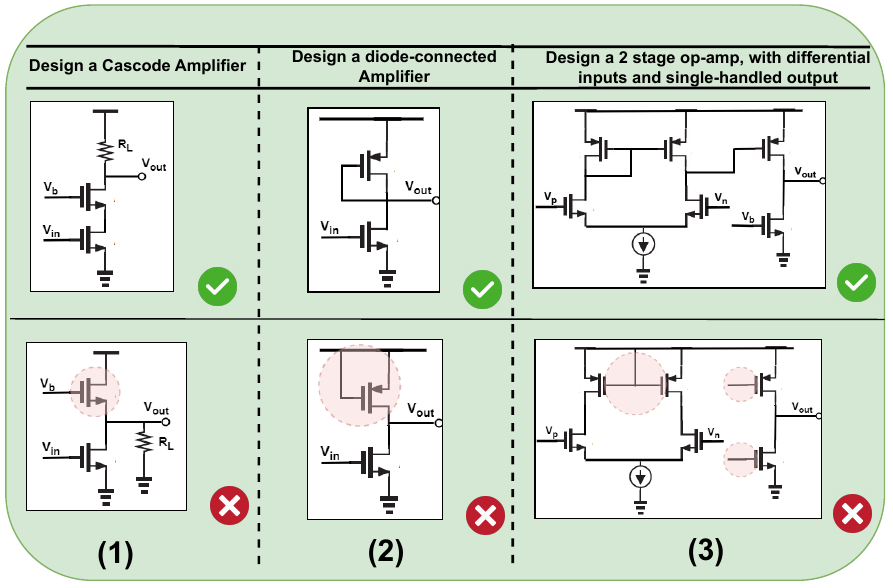}
\vspace*{-0.1in}
\caption{Pass/fail design cases generated by fine-tuned model.}
\label{fig:finetune}
\vspace*{-0.2in}
\end{figure}

~\autoref{fig:finetune} showcases representative examples of both successful and failed netlists generated by LLMs, illustrating critical design challenges and the effectiveness of our approach. Notably, in~\autoref{fig:finetune}: (1) demonstrates the improper use of an NMOS as a current source rather than forming the intended cascode configuration; (2) highlights a fundamental error where the gate of a diode-connected transistor, expected to connect to the drain, is incorrectly connected to the source; (3) the complexity of a two-stage design further amplifies the difficulties in understanding and replicating the intended topology. 
These examples show the limits of unoptimized LLMs. Through the finetuned \texttt{GPT-4o} model, we resolved the issues such as
eliminating floating nets. For more analysis see Appendix~\ref{sec:appendix_c}. 



\subsection{Integration into the AnalogCoder~\cite{lai2024analogcoder} Framework}
AnalogCoder is a feedback-driven LLM framework for automated analog circuit design. \autoref{tab:comp_analog} reports the improvements in Pass@1 and Pass@5 on Tasks 12-20 by using our fine-tuned GPT-4o (\autoref{tab:result}) with AnalogCoder instead of the base GPT-4o model (excluding tasks where AnalogCoder achieves 100\% success). 
AnalogCoder's Pass@1 rate on Task 20 improves from 20\% to 60\% when used with our fine-tuned GPT-4o model. 
The fine-tuned GPT-4o model alone only achieves a Pass@1 of 40\%, suggesting that Masala-CHAI fine-tuning \emph{and} AnalogCoder feedback together provide best performance.

\begin{table}[htbp]
\centering

\caption{Results with AnalogCoder~\cite{lai2024analogcoder}. Many tasks show improvement and for others, results are the same (100\%).}
\label{tab:comp_analog}
\begin{tabular}{ccccc}
\toprule
\multirow{2}{*}{ Task ID } & \multicolumn{2}{c}{AnalogCoder} & \multicolumn{2}{c}{\textbf{AnalogCoder w/ FT \texttt{GPT-4o}}} \\
\cmidrule(lr){2-3} \cmidrule(lr){4-5}
& Pass@1 & Pass@5 & Pass@1 & Pass@5 \\
\midrule
12 & 13.3 & 57.1 & \textbf{80} & \textbf{100} \\
14 & 73.3 & \textbf{100} & \textbf{100} & \textbf{100} \\
15 & 13.3 & 57.1 & \textbf{86.7} & \textbf{100} \\
16 & 6.7 & 33.3 & \textbf{53.3} & \textbf{99.3} \\
17 & 0.0 & 0.0 & \textbf{60} & \textbf{99.8} \\
19 & 60 & 99.8 & \textbf{80} & \textbf{100} \\
20 & 20 & 73.6 & \textbf{60} & \textbf{99.8} \\
\bottomrule

\end{tabular}
\end{table}

\section{Conclusion}
\label{sec:conclude}

Masala-CHAI is an automated framework for creating large-scale SPICE netlist datasets from analog textbooks, resulting in the largest publicly available dataset of ~7500 SPICE netlists with corresponding captions. Fine-tuning LLMs on Masala-CHAI demonstrates its use in automated SPICE netlist generation from natural language prompts. Our dataset and codebase (\url{https://masala-chai-llm.github.io/}) are fully open-sourced and easily extensible on additional textbooks and documentation.




\newpage

\bibliographystyle{IEEEtran}
\bibliography{biblio.bib}

\newpage
  \newpage
   \appendix

\section{Appendix}
 \label{app:case}

\subsection{Evaluation\label{sec:appendix_a}}

\begin{figure*}[h]
\centering
\includegraphics[width=1\textwidth]{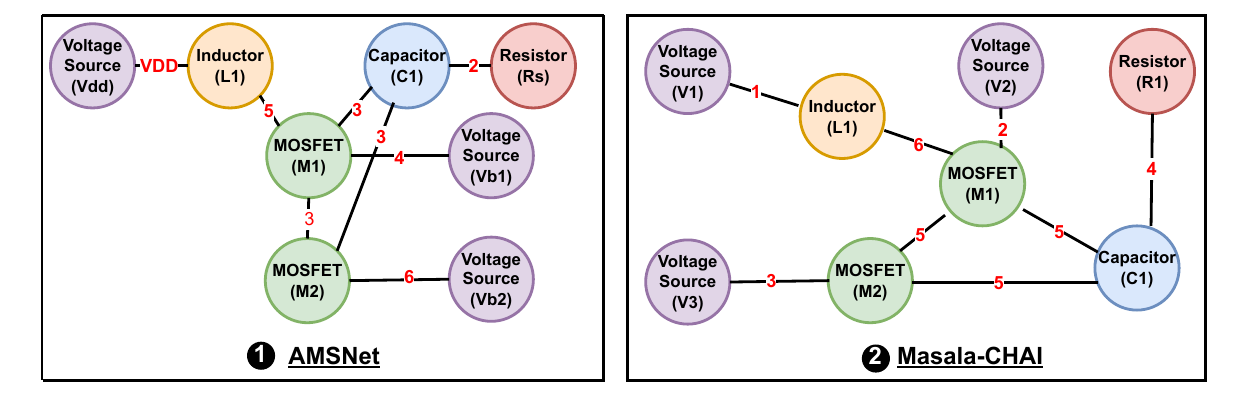}
\caption{Graph representation of SPICE netlists generated by AMSNet~\cite{tao2024amsnet} and Masala-CHAI. They are identical.}
\label{fig:graph}
\end{figure*}
We leverage graph-matching concepts to compare the structural similarity between SPICE netlists. Unlike traditional approaches, we parse the netlists to extract components and their connections. We identify components by types (e.g., \texttt{Resistor}, \texttt{Capacitor}, \texttt{MOSFET}) to ensure that our graph representation is invariant to component labels used during netlist generation. This yields a graph ``G'', where nodes are circuit components and edges are interconnections, as illustrated~\autoref{fig:graph}.
We use Graph Edit Distance (GED) to compare representations of two SPICE netlists. GED ($G_1$,$G_2$) calculates the sum of the costs associated with transforming $G_1$ into $G_2$ through a series of edit operations on nodes and edges, such as insert, delete, and substitute. 
We normalize the GED to yield a final metric in the range 0 to 100. This normalization is achieved by calculating the maximum GED, obtained by summing the number of nodes and edges in both graphs and scaling accordingly. This metric allows for a consistent comparison across different circuits. Thus, the similarity score between the two netlists is: 
\begin{equation}
S = \left( 1 - \frac{\text{GED}(G_1, G_2)}{\text{GED}_{\text{max}}} \right) \times 100\%
\end{equation}

\subsection{Verifying against AMSNet~\cite{tao2024amsnet} netlists\label{sec:appendix_b}}

We demonstrated the effectiveness of our approach by evaluating it using the AMSNet~\cite{tao2024amsnet} dataset. Unlike AMSNet, which relies on an algorithmic method supplemented with some manual effort to generate SPICE netlists, our method introduces a fully automated flow (\autoref{fig:flow}). When comparing the SPICE netlists generated by our method to those produced by AMSNet, our approach achieved a \textbf{100\%} similarity score, as defined earlier. This validates the robustness of our flow and motivates us to extend the evaluation to additional circuit diagrams.~\autoref{fig:graph} illustrates an example 
of graphs representation of SPICE netlist generated by AMSNet (\circleone{}) and our \textbf{Masala-CHAI} framework (\circletwo{}).
 The results demonstrate that the number of components and their neighboring elements is identical across both methods, with differences in component and net naming conventions. These findings strengthen Masala-CHAI accuracy and scalability in generating accurate netlists for various circuit diagrams.


\subsection{Role of high-quality dataset on Finetuning\label{sec:appendix_c}}

\begin{figure*}[ht]
\centering
\includegraphics[width=\textwidth]{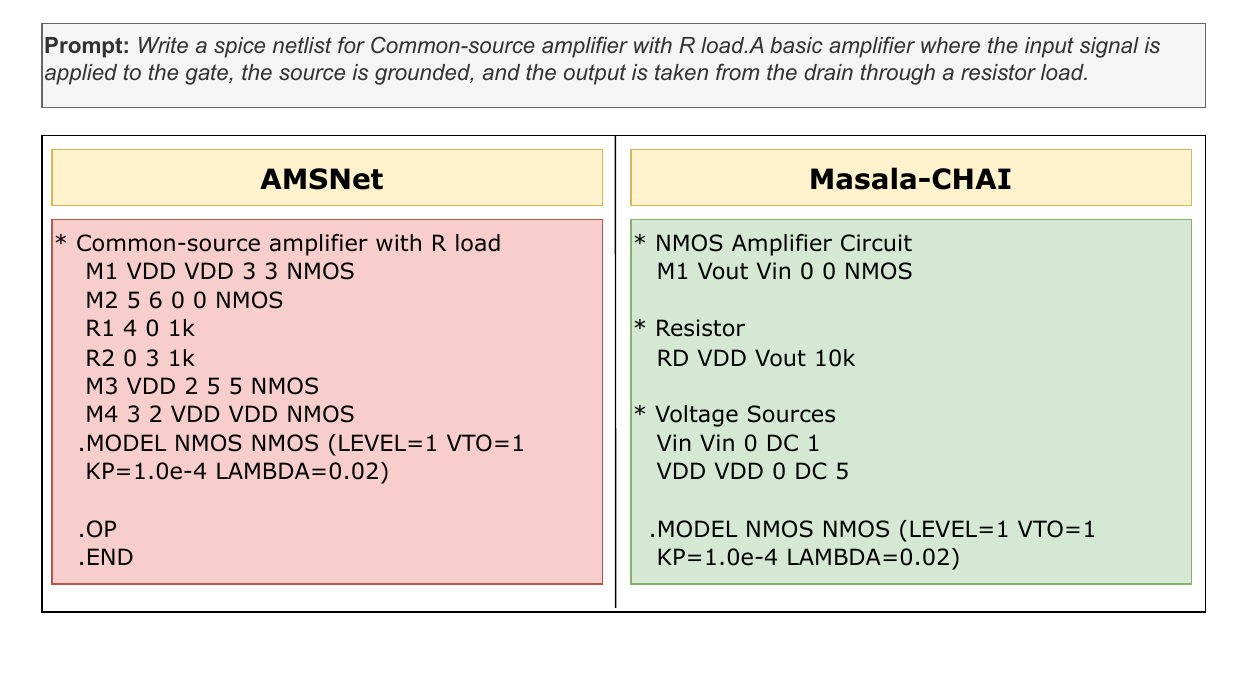}
\caption{Example showing responses from a fine-tuned \texttt{GPT-4o} model with AMSNet and MASALA-CHAI dataset, respectively.}
\label{fig:amsnet}

\end{figure*}

As a part of our study, we fine-tuned the \texttt{GPT-4o} with the AMSNet~\cite{tao2024amsnet} dataset as well and compared the quality of response with the MASALA-CHAI dataset.~\autoref{fig:amsnet} shows the responses for a simple \textit{``Common Source Amplifier''} (ID=1) from the fine-tuned model using AMSNet and MASALA-CHAI dataset, respectively. AMSNet lacks a description of the SPICE netlist present in the dataset, which hinders its performance for supervised fine-tuning (SFT), making it not ready to use for LLM-based tasks that require high-quality description alongside the code for unlocking its full potential. MASALA-CHAI also offers such a pipeline to generate description alongside the SPICE netlist from a document, as shown~\autoref{fig:flow} (\circleeight{}).


\end{document}